\documentstyle[prb,aps]{revtex}
\begin{document}

\title{Kondo Effect in Fermi Systems with a Gap: A Renormalization Group Study}

\author{Kan   Chen$^{(1)}$ and   C. Jayaprakash$^{(2)}$}

\address{ $^{(1)}$Department  of  Computational Science,
National University  of Singapore, Singapore 119260\\
and\\
$^{(2)}$Department  of  Physics, The Ohio State University,
Columbus, OH 43210, U.S.A.}

\date{\today}

\maketitle
\begin{abstract}
We  present the results of a Wilson Renormalization Group study of the 
single-impurity Kondo and Anderson models in a system with a gap in the
conduction electron spectrum. The behavior of the 
impurity susceptibility and the zero-frequency response
function, $T<<S_z; S_z>>$ are discussed in the cases with and without
particle-hole symmetry. In addition, for the asymmetric Anderson model 
the correlation functions, 
$<\vec S \cdot\vec \sigma (0)>$, 
$<n_d>$, and $<n_d(2-n_d)>$ are computed. 

\vspace{4pt}
\end{abstract}

\section{Introduction}

 The properties of  a magnetic impurity in a semiconductor or an insulator 
are 
of interest for a variety of reasons. In a normal fermi system a  
spin-$\frac{1}{2}$ impurity yields logarithmic temperature dependences in the
impurity susceptibility and the resistivity at high temperatures; 
at low temperatures the magnetic moment is 
quenched \cite{kgw,nozieres}. The existence of a sharp fermi surface and the 
concomitant occurrence
of (low-energy) particle-hole pairs play an important role in understanding
the behavior of the model. Thus it is interesting from a theoretical point of
view to understand whether the
Kondo effect persists and  under what conditions 
quenching occurs in a system with a gap and 
determine the behavior quantitatively. We also note that 
the Anderson
impurity has been studied in the
context of the logarithmic temperature dependence of the conductivity of 
trans-polyacetylene \cite{cruz}; the system was modeled by a 
continuum Hamiltonian 
%proposed by Takayama {\em et al.} \cite{taka}
that exhibits a gap due to Peierls
distortion. The impurity model was investigated using a Hartree-Fock
closure of 
the equation of 
motion. In addition, a variety of Kondo and valence fluctuating 
insulators (modeled theoretically by a Kondo or Anderson lattice) 
such as $SmB_6$ and 
$Ce_3Bi_4Pt_3$ among others provide another motivation for studying the
single impurity problem in a system with a gap.

In this paper we present the results of our study of the Kondo and Anderson 
impurities in a system with  a gap.
We  apply Wilson's (numerical) Renormalization Group (RG) technique using a 
variant of a numerical tridiagonalization method
devised by us earlier\cite{cj1} and provide results for both the
susceptibility
and zero-frequency response functions. We also discuss a simple effective
Hamiltonian that allows us to understand the physics underlying our results. 

We begin with a summary of previous work on the problem; in the next 
section we provide a
brief sketch of the technique that is described in detail in the literature
\cite{kww}. We then present the results of our numerical simulations and in
the
final section we discuss the effective Hamiltonian description of our results.

%The problem of magnetic impurities in pseudo-gap Fermi systems was first
%investigated by Withoff and Fradkin \cite{Withoff}, who studied 
%the Kondo model with the density of states $\rho(\epsilon)=C|\epsilon|^r$
%(the Fermi energy is set to zero)
%using perturbative scaling and the $1/N$ expansion to leading order. 
 
We begin with a brief overview of previous work. The first calculations 
were done by Ogura
and Saso \cite{saso0,saso1,ogura1} and Takegahara {\em et
al.} \cite{take1,take2}.
Ogura and Saso used a $1/N$ expansion of the degenerate Anderson model and
found to leading order a transition between the triplet and singlet ground 
states when the gap $E_g$ equals twice the Kondo temperature $T_K$. 
In their Quantum Monte Carlo 
(QMC) simulations they found
indications of a similar transition even for the symmetric Anderson model
\cite{saso0}; for the asymmetric model they obtained a transition 
between the different 
ground states at approximately $E_g \approx 3T_K$.
We note that their QMC computations were limited to temperatures above
$T_K/10$.

Takegahara, Shimizu, and Sakai \cite{take1,take2} used both
Quantum Monte Carlo simulations and the numerical Renormalization group method
of Wilson. They considered the symmetric Anderson model 
and found that at low temperatures the susceptibility follows a Curie law
resulting from an unquenched magnetic moment. 
They observed the crucial difference 
between symmetric and asymmetric Anderson models cases; when particle-hole
symmetry is obeyed the moment remains unquenched for all non-zero values 
of the gap while there is a transition
in the asymmetric case. They also used the Wilson numerical Renormalization
group to follow the spectrum of the low-energy states but not the
susceptibility. It is difficult to use their version of the numerical RG 
formulation to calculate low temperature (much less than the band gap) 
properties of the model.

Yu and Guerrero \cite{yu} studied an Anderson impurity
in a semiconducting host using the density matrix renormalization technique.
Their calculation which is restricted to $T=0$ considered electron
spin-impurity
spin correlation functions and 
found no {\em qualitative
difference} between the symmetric and asymmetric cases. We will comment on
this
point later. 
The importance of the particle-hole symmetry breaking has been 
emphasized recently \cite{inger} in the context of the 
Kondo problem with a pseudogap. 
In this work the impurity
susceptibility for the case of a Kondo system with a gap was also calculated
using the Wilson renormalization group method: 
in the particle-hole symmetric case 
the impurity retains its moment in the ground state
for all $J$; in the presence of potential scattering 
the moment is completely quenched
provided that $\Delta<<T_K$.  These results are in agreement with those of 
Takegahara {\em et al} \cite{take1,take2}.

We have made a comprehensive study of the Kondo and Anderson models in 
gapped systems with and
without particle-hole symmetry breaking using the numerical RG method. 
Our RG formulation is based on the numerical tridiagonalization technique
developed by us \cite{cj1}, which allows us to calculate various quantities
in the entire temperature range. We
report results for a zero-frequency response function, correlation
functions and
the susceptibility: we emphasize the differences in their behaviors in the 
various
regimes and clarify which of these are good probes of the nature of the
low-temperature fixed point behavior. 

%\pagebreak

\section{Wilson's RG Formulation}

We consider the Kondo and Anderson models with a conduction electron 
Hamiltonian with the density of states $\rho(\epsilon)$; as a function of the
energy $\epsilon$, $\rho$ is a constant for
$D_0 > |\epsilon| > \Delta_0$, where the band edges lie at $\pm D_0$ 
from the Fermi level which is chosen to be in the middle of the gap. 
The width of the gap is thus $2\Delta_0$. The impurity part of the
Hamiltonian  for the spin$-\frac{1}{2}$ Kondo
problem with the impurity spin denoted by $\vec S$ is given in standard
notation by
\begin{eqnarray}
\label{kondo}
H_K &=& -J 
\sum_{\mu,\nu}~\int_{-D_0}^{D_0}d\epsilon\sqrt{\rho(\epsilon)}~ 
\int_{-D_0}^{D_0}d\epsilon'\sqrt{\rho(\epsilon')}~~\vec{S}~
\cdot ~c^+_{\epsilon,\mu}~\frac{1}{2}
\vec{\sigma}_{\mu\nu} ~c_{\epsilon',\nu} \nonumber \\
& &+K\sum_{\mu}~\int_{-D_0}^{D_0}d\epsilon\sqrt{\rho(\epsilon)}
\int_{-D_0}^{D_0}d\epsilon'\sqrt{\rho(\epsilon')}~~c^+_{\epsilon,\mu}
c_{\epsilon',\mu}  ~~~;
\end{eqnarray}
for the Anderson model  we have 
\begin{eqnarray}
\label{anderson}
H_A &=& (\epsilon_d ~+~ \frac{U}{2})\sum_{\mu}~
c^+_{d\mu}c_{d\mu}~+~\frac{U}{2}\left (\sum_{\mu}~
c^+_{d\mu}c_{d\mu}-1\right)^2  \nonumber \\
& &+\sum_{\mu}~\int_{-D_0}^{D_0}d\epsilon\sqrt{\rho(\epsilon)}~~ 
[Vc^+_{\epsilon,\mu}c_{d\mu}~+~ V^*c^+_{d\mu}c_{\epsilon,\mu}] ~~~,
\end{eqnarray}
where $c^+_{d\mu}$ creates an electron with spin $\mu$ at the impurity placed
at the origin. The choice $K=0$ in the Kondo problem and
$\epsilon_d+\frac{U}{2}=0$ in the Anderson model correspond to particle-hole
symmetry. \\

Following Wilson we perform a  {\em logarithmic}
discretization of the energy variable; we rescale the energy by $D_0$ so that 
$\epsilon \in [-1,1]$, introduce a scale factor 
$\Lambda(>1$), and define 
 the $n$th interval for positive $\epsilon$ to lie between  
$\Lambda^{-n-1}$ and 
$\Lambda^{-n}$. The band gap is chosen to be 
$$\Delta = \Lambda^{-M_0} ~~.$$
Next we replace the continuous set of energy levels in 
the $n$th interval  [$\Lambda^{-n-1}$, $\Lambda^{-n}$]
and [$-\Lambda^{-n}$, $-\Lambda^{-n-1}$] by single levels 
at ($\Lambda^{-n-1}$+$\Lambda^{-n}$)/2 and $-(\Lambda^{-n}$+$
\Lambda^{-n-1}$)/2 respectively, and introduce
$a^+_{n\mu}$, $b^+_{n\mu}$, the conduction electron creation operators 
for the states with the corresponding 
energies  ($\Lambda^{-n-1}$+$\Lambda^{-n}$)/2 and 
$-(\Lambda^{-n-1}+\Lambda^{-n})/2$. 
After this discretization, the Anderson Hamiltonian can be rewritten in the
following form \cite{kww} (Here we only consider the Anderson Hamiltonian; 
similar RG formulation can be written down for the Kondo Hamiltonian)
\begin{eqnarray}
H_A &=& \frac{1+\Lambda^{-1}}{2}\sum_{m=0}^{M_0-1}\Lambda^{-m}(a^+_{m\mu} 
a_{m\mu}-b^+_{m\mu}
b_{m\mu})~+~(\epsilon_d+\frac{1}{2}U)~c^+_{d\mu}c_{d\mu} \nonumber
\\
& & ~+~ \left[\frac{2\Gamma}{\pi}\right]^{\frac{1}{2}}~(f^+_{0\mu}c_{d\mu}+
c^+_{d\mu}f_{0\mu})~+~\frac{U}{2}(c^+_{d\mu}c_{d\mu}-1)^2 ~~~,
\end{eqnarray}
where 
$$f_{0\mu} = \sqrt{\frac{1-\Lambda^{-1}}{2(1-\Delta)}}
\sum_{m=0}^{M_0-1}\Lambda^{-m/2}(a_{m\mu}+b_{m\mu}) ~~~.$$
The initial values of the couplings $\Gamma$ ($\equiv 0.5\pi V^2$),
 $\epsilon_d$, and
$U$, are now in units of $D_0$ (taken to be one); the gap in the density of
states 
is between $\Delta$ and $-\Delta$. 

We use the (numerical) tridiagonalization scheme devised by us 
earlier \cite{cj1} to transform 
the Hamiltonian
to the following tridiagonalized form:
\begin{eqnarray}
H_A &=& \frac{1+\Lambda^{-1}}{2}\sum^{N_0 -1}_{n=0}[\xi_{n}
	(f^{+}_{n\mu}f_{n+1\mu}+ h.c.)]
~+~(\epsilon_d+\frac{1}{2}U)~c^+_{d\mu}c_{d\mu} \nonumber  \\
& & ~+~ \left[\frac{2\Gamma}{\pi}\right]^{\frac{1}{2}}~(f^+_{0\mu}c_{d\mu}+
c^+_{d\mu}f_{0\mu})~+~\frac{U}{2}(c^+_{d\mu}c_{d\mu}-1)^2,
\end{eqnarray}
where $N_0 = 2M_0-1$.

In order to carry out the RG calculation, we need to rescale the Hamiltonian
at each iteration step. The rescaling is done by defining $H_N$ as
follows:
\begin{eqnarray}
H_N &=&\frac{1}{\xi_{N-1}}\left[~\sum^{N -1}_{n=0}
[\xi_{n} (f^{+}_{n\mu}f_{n+1\mu}+h.c.)] ~\right] \nonumber \\
& & +
~\frac{1}{\xi_{N-1}}\left[(\tilde{\epsilon}_d+\tilde{U})~c^+_{d\mu}c_{d\mu} 
+\tilde{\Gamma}^{\frac{1}{2}}~(f^+_{0\mu}c_{d\mu}+
c^+_{d\mu}f_{0\mu})~+~\tilde{U}(c^+_{d\mu}c_{d\mu}-1)^2 \right], \nonumber\\
\end{eqnarray}
where $\tilde{\epsilon}=\frac{2}{1+\Lambda^{-1}}\epsilon_d$, 
$\tilde{U}=\frac{U}{1+\Lambda^{-1}}$,
$\tilde{\Gamma}=[\frac{2}{1+\Lambda^{-1}}]^2\frac{2\Gamma}{\pi}$, and 
the rescaling factor is $S_N = \frac{2}{(1+\Lambda^{-1})\xi_{N-1}}$.

The recursion relation can be written in the following compact form: 
$$H_{N+1} = \frac{\xi_{N-1}}{\xi_N}H_N +(f^{+}_{N\mu}f_{N+1\mu} ~+~h.c.).$$
This recursion relation enables one to set up an iterative diagonalization
scheme to calculate the energy levels of $H_N$ and thus to determine
thermodynamic properties; the  recursion is implemented numerically and 
is stopped 
at $N=N_0$ corresponding to the edge of the gap below which there are no
conduction electron states.
Recall that as we increase $N$, the system effectively evolves from high 
temperatures to 
low temperatures. At a given $N$, the thermodynamic quantities are 
calculated for $T_N=1/(\bar{\beta}S_N)$ for a selected values of 
$\bar{\beta}$. By studying the evolution of the many-body energy level 
structures we also obtain 
information near the fixed points of the Hamiltonian. 

For $N<N_0$, the thermodynamic quantities are calculated for $T_N =
1/(\bar{\beta} S_N)$ for a selected value of $\bar{\beta}$; the accuracy of
the
numerical evaluations is enhanced by performing a second-order
perturbation calculation by writing the Hamiltonian as
$$H_A=(H_N+H_{I}+H_B)/S_N,$$
where 
	$$H_I = \frac{\xi_N}{\xi_{N-1}}(f^{+}_{N\mu}f_{N+1\mu} +~h.c.),$$
and
	$$H_B = \frac{1}{\xi_{N-1}}\left\{\sum^{N_0 -1}_{n=N+1}[\xi_{n}
	(f^{+}_{n\mu}f_{n+1\mu}+~h.c.)]\right\}.$$

For $N=N_0$, the thermodynamic quantities are calculated for a
sequence of temperatures $\{T_l\}$. 
Since $H_{N_0}$ is the full
Hamiltonian (hence, no second order perturbation is needed), we can calculate
the quantities at temperatures much lower than typical energy scale at
$N=N_0$, which is the bandgap $\Delta$. We choose $T_l$
to be a sequence of values from $0.175$ of the maximum energy kept in
the many body states of $H_{N0}$ to $0.000175$ of the maximum energy.
Thus the thermodynamic quantities at low temperatures are calculated with
the ``effective Hamiltonian'' $H_{N_0}$. 

%\pagebreak

\section{Results}

We present the results obtained from our numerical calculations for the two
models. 

\subsection{Kondo Model}

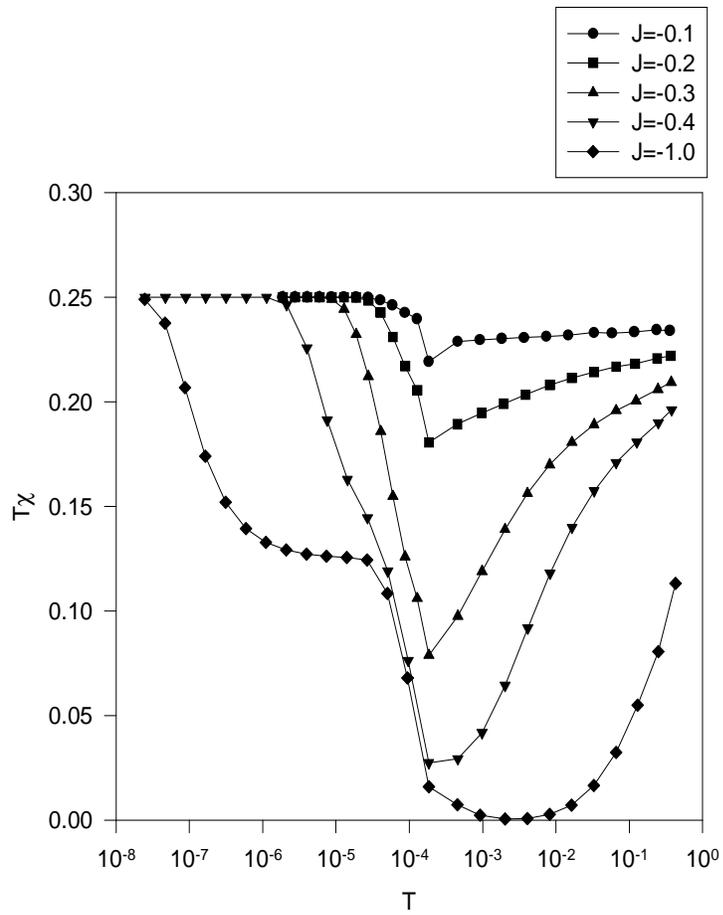
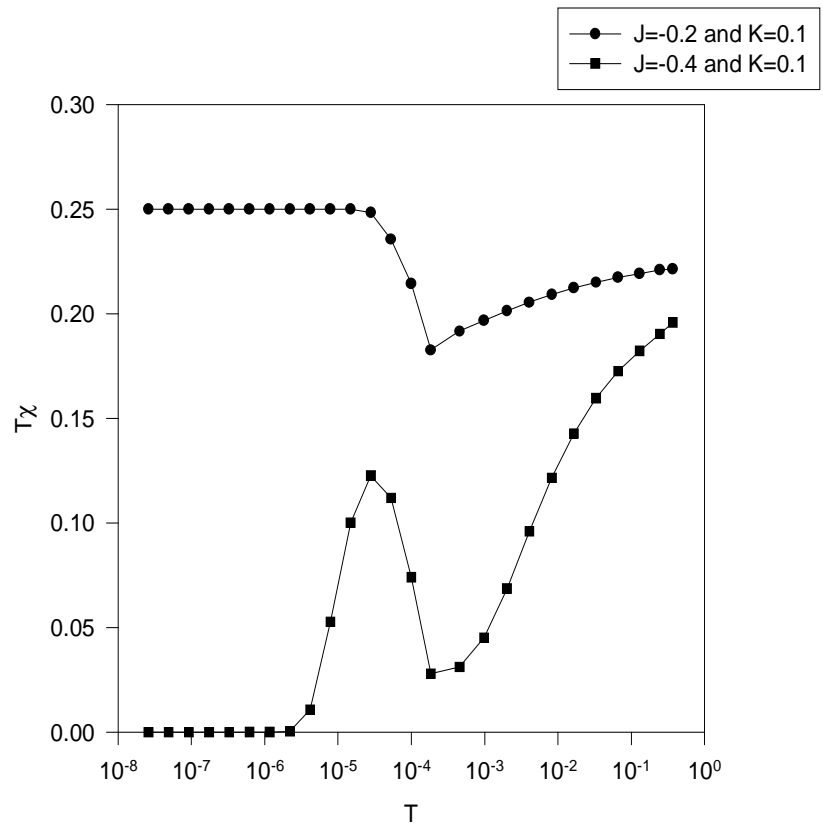
\begin{figure}
\caption{$T\chi_{imp}$ plotted as a function of $T$ for the Kondo
problem.  The gap energy is $\Delta= 1.22\times 10^{-4}$.
(a) The potential scattering is absent ($K=0.0$). 
The values of the coupling $J$ used are $-0.1, -0.2, -0.3, -0.4$, and 
$-1.0$. Note that as $T \rightarrow 0$, 
$T\chi$ approaches $1/4$. (b) Particle-hole symmetry breaking is present
($K=0.1$). Note
that for $J= -0.4$, $T\chi$ goes to zero, while for $J=-0.2$, 
it approaches $1/4$.
}
\label{chi_kondo}
\end{figure}

Our calculations were performed using a scale factor of 
$\Lambda = 2$ and a band gap energy $\Delta = 1.22\times 10^{-4}$ 
corresponding to $M_0 = 13$ .
The first and obvious quantity to
consider is the impurity susceptibility, $\chi$, which we emphasize is 
defined as the total susceptibility of the system minus the
susceptibility of the pure system.
In the Kondo problem in the absence of potential scattering 
$T\chi$ approaches the value $1/4$ as $T \rightarrow 0$ for {\em any}
finite bandgap.  The ground state is a magnetic doublet,
its quantum numbers are $(Q=0,~ S=1/2)$.
This is  in
agreement with the results of Takegahara et al \cite{take2} 
for the symmetric Anderson model. 
 The susceptibility curves are displayed in
 Fig.~\ref{chi_kondo}(a). 
Note that some data obtained at intermediate points have 
been suppressed for clarity in this figure as well as in other
figures we are going to present in this paper. 
The calculation is done for initial values of the coupling 
given by $J_0 = -0.1,-0.2,-0.3,
-0.4,-1.0$. 
Note that for large values of $|J|$ the universal shape of $T\chi$ of 
the ordinary
Kondo problem is evident at high temperatures but below temperatures of the
order of the gap $T\chi$ increases sharply. 

The effect of particle-hole symmetry breaking introduced by 
potential scattering is very
important as has been noted before \cite{inger,take1}.
  The results for $T\chi$ are displayed in Fig.~\ref{chi_kondo}(b). 
For $K_0=0.1$ and $J_0=-0.2$ ($T_K \approx 7.4 \times 10^{-6}<<\Delta$),
$T\chi$ again goes to 1/4 as $T$ goes to zero.
For stronger Kondo coupling, $J=-0.4$ ($T_K \approx 2.1 \times
 10^{-3}>>\Delta$), the impurity spin is quenched and
$T\chi \rightarrow 0$. There is a discontinuous (``first-order'') 
transition due to a crossing of energy levels. Crudely speaking, 
in the generic case without particle-hole symmetry, the transition 
occurs when the energy gained by forming 
the singlet which is of the order of $T_K$ is larger than the energy 
required to create a particle-hole excitation across the gap. 

We have also calculated the zero-frequency response function
$T<<S_z; S_z>>$. The techniques for performing such calculations have been
explained in an earlier paper for the ordinary Kondo problem \cite{cjk}. 
We used $\Lambda=3.0$ in the calculation of the response
function. Note that for the ordinary Kondo problem $<<S_z; S_z>>$
is essentially the same as the impurity susceptibility $\chi$ for
 small values of the initial coupling \cite{cjk}. However, for the 
density of
states with a gap, $<<S_z; S_z>>$ and $\chi$ behave quite differently
at low temperatures (when $T<\Delta$). 
In the absence of potential scattering, in 
contrast to $T\chi$ which
approaches a fixed value of $1/4$ as $T \rightarrow 0$, 
$T<<S_z; S_z>>$ approaches a value $C_0$ 
which depends on the bandgap; this persists also when the moment is not 
quenched in the presence of potential scattering. The results for
$C_0$ are listed in Table I. For $K=0$ our results are consistent with 
the value $C_0$ being proportional to $\Delta^2$ for
$\Delta <<T_K$. We will derive this 
result from our effective Hamiltonian description in the next section. 
This result agrees with the claim made by Takegahara 
et al. for the susceptibility \cite{take1}; we note that 
they appear to have 
identified 
$<<S_z; S_z>>$ with the impurity susceptibility. 
In the presence of potential scattering when $\Delta$ is increased 
for fixed $T_K$ the ground state changes abruptly from a singlet 
[$(Q=-1, S=0)$] to a doublet. The value of $T\chi$ jumps from
$0$ to $1/4$ and correspondingly the value of $C_0$  also jumps
discontinuously.

\subsection{Anderson Model}

The calculations for the Anderson model were performed with  
the parameter $\Lambda=3$. A range of values was used for the band gap 
$\Delta=\Lambda^{-M_0}$: the value of $M_0$ was varied between $3$ and $19$.

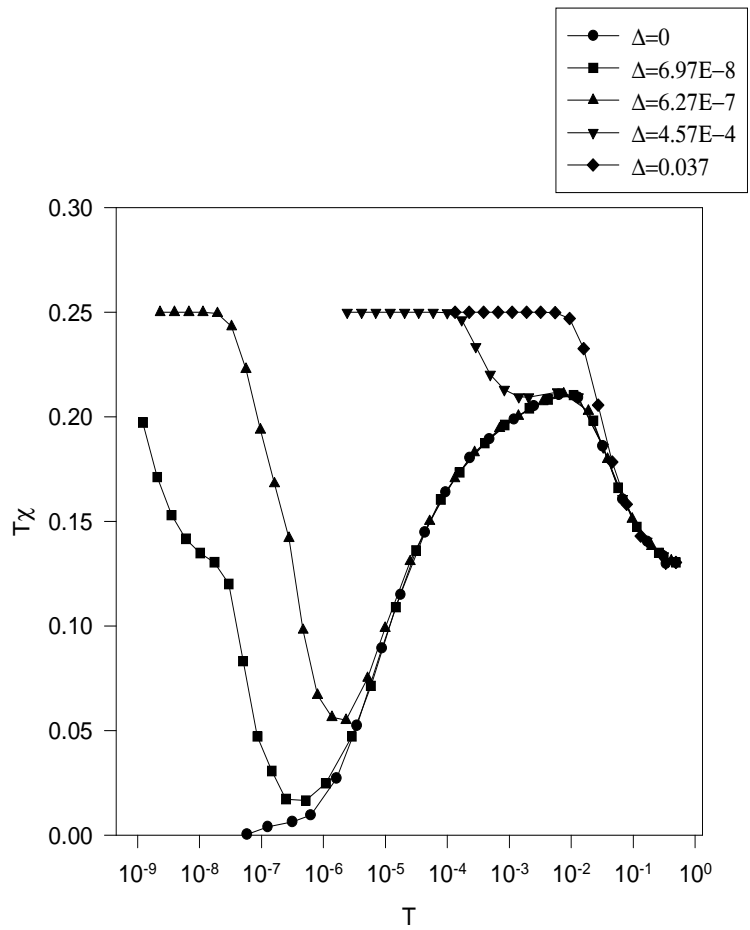
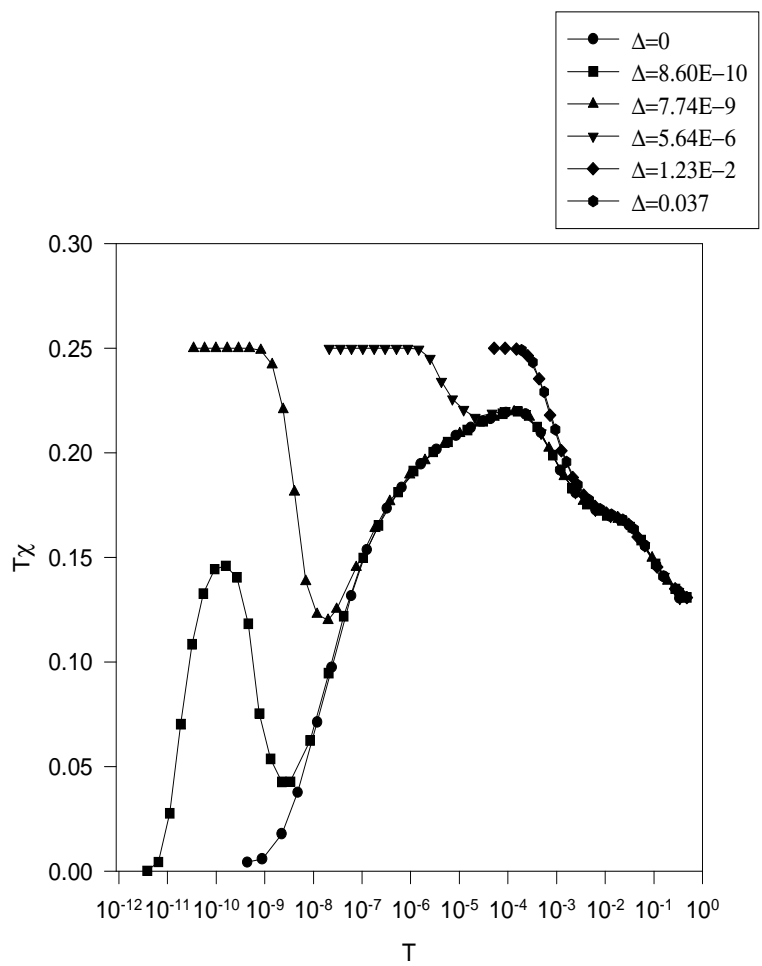
\begin{figure}
\caption{$T\chi_{imp}$ plotted as a function of $T$ for 
the Anderson model. (a) Symmetric Anderson model
with $U=0.1, \epsilon_d=-0.05$, and $\Gamma=0.006$. Note that as $T \rightarrow 0$, $T\chi$ approaches $1/4$. 
(b) Asymmetric Anderson model with $U=0.1, \epsilon_d=-0.001$,
and $\Gamma=0.00015$. Note that as $T \rightarrow 0$, $T\chi$ approaches $1/4$ if $\Delta>>T_K$ and $0$ if $\Delta <<T_K$.
The values for the gap energy $\Delta$ used in
the calculations are shown in the legends of the figures.
}
\label{chi_anderson}
\end{figure}

For the symmetric Anderson Model, with $U=0.1$,
$\epsilon_d=-\frac{U}{2}=-0.05$,
and $\Gamma=0.006$,  $T\chi$ reaches the value of $1/4$ as zero temperature is
approached irrespective of 
the value of the bandgap $\Delta$, signaling a doublet ground state and an
unquenched impurity moment.  If 
$\Delta<< T_K\approx 5.12\times 10^{-6}$, $T\chi$ first decreases toward 
zero along the universal Kondo curve; however, when $T<\Delta$, it rises 
to 1/4 as T goes to zero. 
If $\Delta$ is comparable or larger than $T_K$, on the other hand, $T\chi$
gradually increases to 1/4. Our results for $T\chi$ are displayed in 
Fig.~\ref{chi_anderson}(a). 

The case of the the asymmetric Anderson Model was studied using the parameter
values $U=0.1$, $\epsilon_d=
-0.0001$, and $\Gamma=0.00015$, and the results are displayed in
 Fig.~\ref{chi_anderson}(b). When $\Delta=0$, the system goes successively 
through the free-orbital regime, the mixed valence regime, the 
local moment regime,
and the frozen moment regime \cite{kww}. When $\Delta \neq 0$, for 
$T> \Delta$, the
$T\chi$ curve initially follows the curve for $\Delta=0$ as the temperature
is lowered. When $T$ drops below $\Delta$, $T\chi$ curves starts to rise. 
For $\Delta>T_K$,
the curve continues to rise to 1/4 as $T$ goes to zero. On the
other hand, when $\Delta<T_K$, the curve stops rising, turns over and 
tends to zero
as $T \rightarrow 0$. This behavior is clearly similar to that of 
the $K\neq 0$ case of the 
Kondo model.

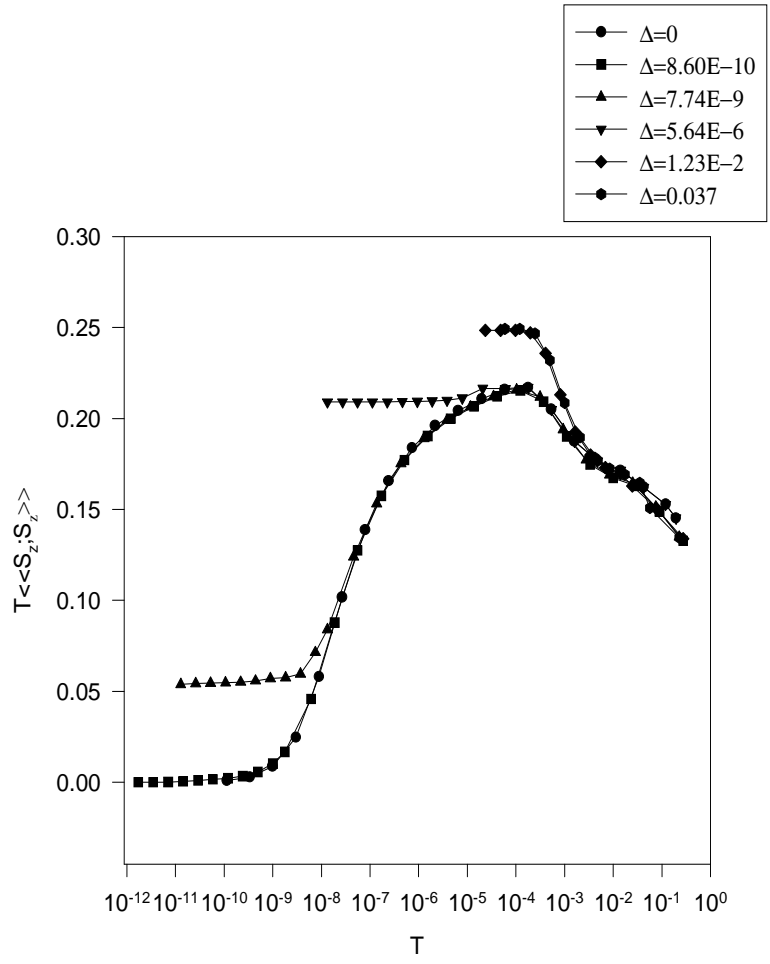
\begin{figure}
\caption{The zero-frequency response function
$T<<S_z; S_z>>$  plotted as a function of $T$ for 
the asymmetric Anderson model with $U=0.1, \epsilon_d=-0.001$,
and $\Gamma=0.00015$. Note the qualitative differences between 
$T<<S_z; S_z>>$ and $T\chi$ at low temperatures.
}
\label{sz_sz}
\end{figure}

Fig.~\ref{sz_sz} shows the general temperature dependences of the
zero-frequency response function $T<<S_z; S_z>>$ for the asymmetric
case.  There is  no qualitative difference in the behavior of 
the response function $T<<S_z; S_z>>$ between the 
Kondo and Anderson models. In the symmetric case where the ground state is
characterized by $(Q=0,S=1/2)$, we again found that the zero-temperature 
value $C_0$ is proportional to $\Delta^2$ when $\Delta$ decreases. 
Also for the asymmetric case $C_0$ jumps discontinuously as the ground state 
changes from a singlet to a doublet as $\Delta$ is increased.

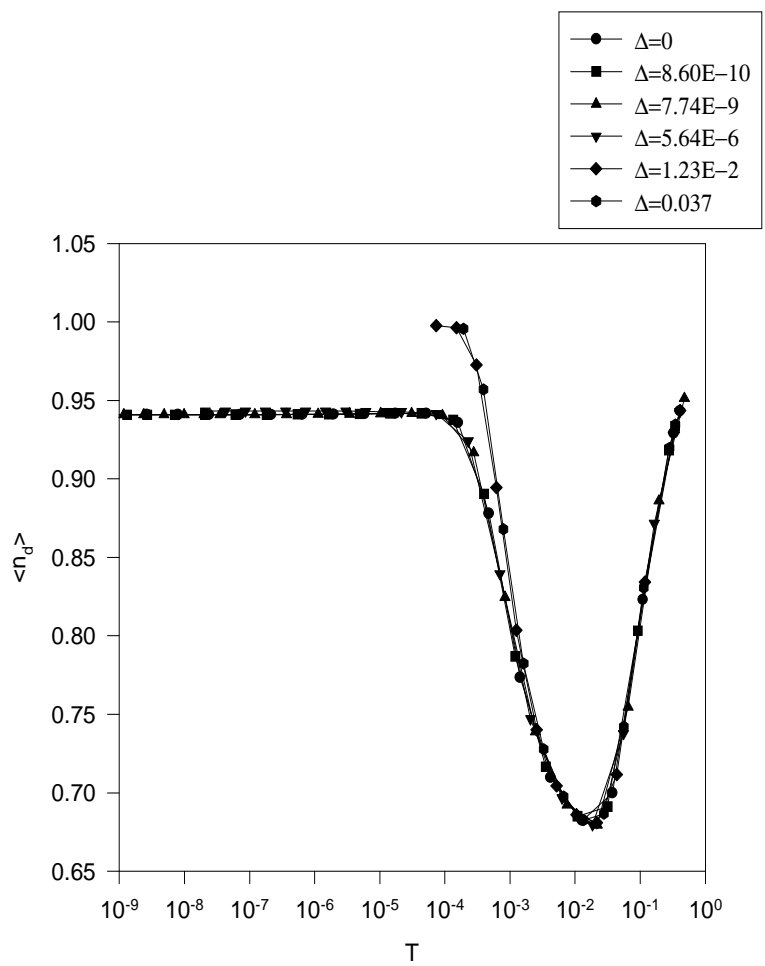
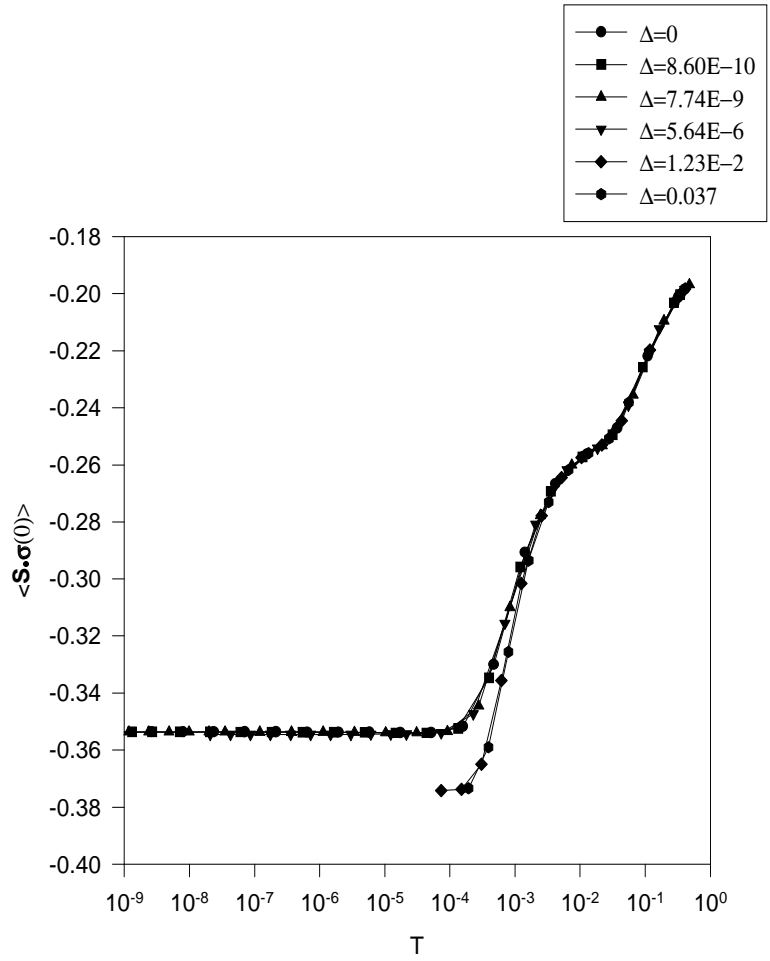
\begin{figure}
\caption{The local correlation functions: (a) $<n_d>$ and
(b) $<\vec S\cdot \vec \sigma(0)>$, plotted as a function 
of $T$ for  the asymmetric Anderson model with $U=0.1, \epsilon_d=-0.001$,
and $\Gamma=0.00015$. Note that when $\Delta<<T_K$, the
correlation functions approach constant values independent of
$\Delta$ as $T \rightarrow 0$.}
\label{corr_func}
\end{figure}

In addition, we have also computed the following correlation functions: 
$<\vec S \cdot \vec \sigma (0)>$, $<n_d>$, and $<n_d(2-n_d)>$.  
Representative figures are shown in Fig.~\ref{corr_func}(a),
and Fig.~\ref{corr_func}(b). Here the main point to be emphasized is that 
once there are no charge fluctuations (for example, when the system 
approaches the local moment regime, or when $T<\Delta$)
the correlation functions 
do not change and approach constant values. In particular, when the
local moment regime is reached ($\Delta$ is less than
the temperature for the local moment formation), the correlation 
functions tend to the same constants as $T\rightarrow 0$ independent of 
the band gap. While the mixed valence regime is 
still reflected in the 
temperature dependence of the correlation functions, the Kondo 
effect does not show up in the correlation functions.
 This point is not very well appreciated. One simply {\em cannot} investigate
the Kondo effect using {\em local} correlation functions, such as the 
impurity spin-conduction electron spin-density at the origin, since they do
not
contain information about the system on the energy scale of $T_K$ or 
equivalently the length scale
of $\hbar v_e/T_K$ where $v_e$ is the characteristic velocity of the
electrons. We believe that this is the reason why in the work of
Yu and Guerrero \cite{yu} no difference was found
between the symmetric and asymmetric Anderson models in the correlation
functions at short length scales.

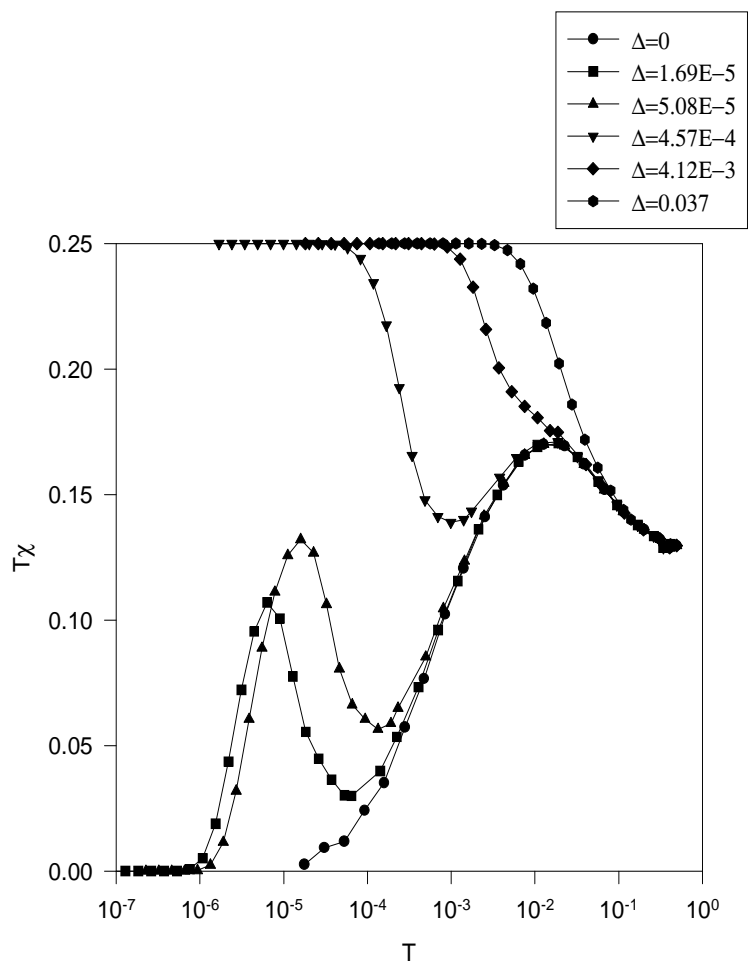
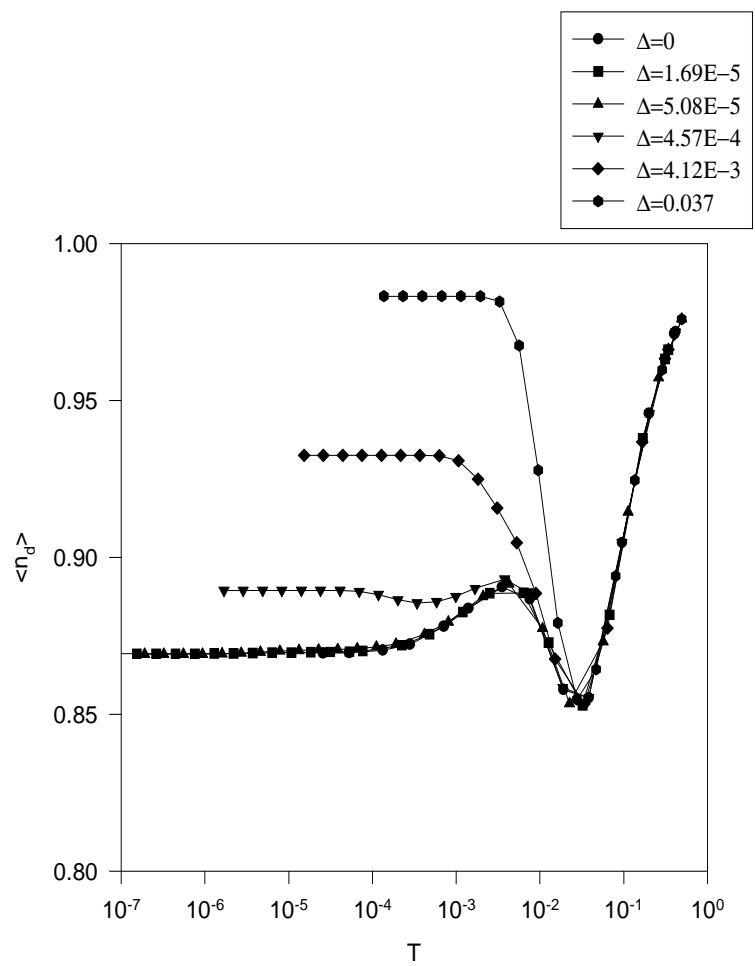
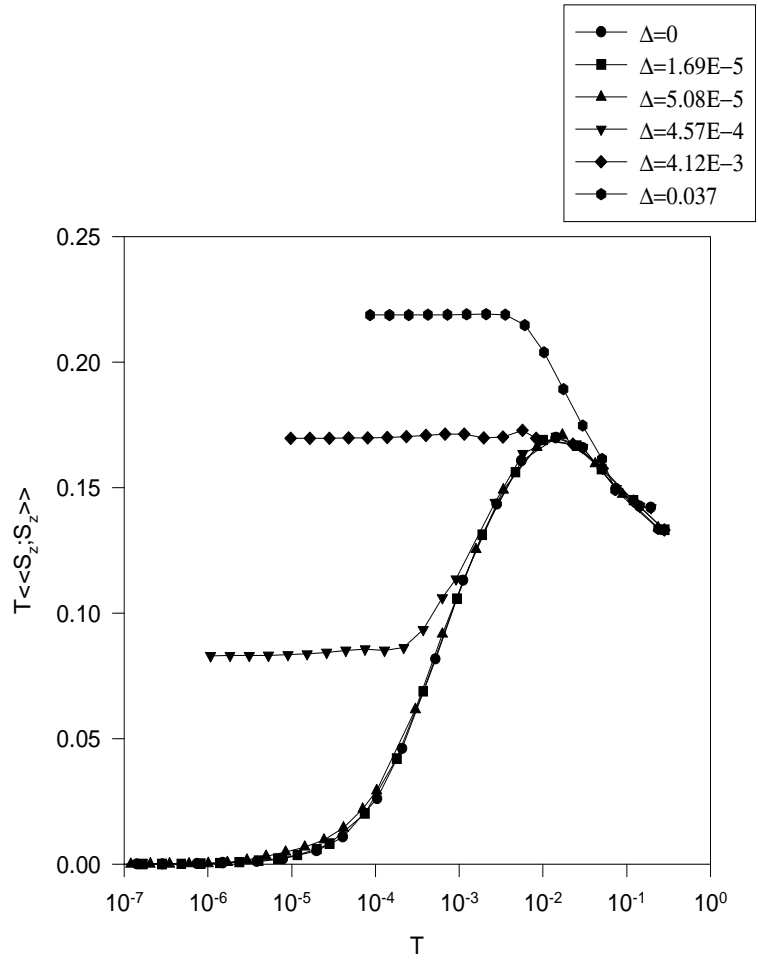
\begin{figure}
\caption{The impurity susceptibility (Fig.~5(a)), $<n_d>$ 
(Fig.~5(b)), and $T<<S_z; S_z>>$ (Fig.~5(c)), plotted as a function 
of $T$ for  the asymmetric Anderson model with $U=0.1, \epsilon_d=-0.025$,
and $\Gamma=0.01$. 
}
\label{mv}
\end{figure}

Finally, we present our results for the mixed-valent regime. We considered
the asymmetric Anderson Model, with $U=0.1$, $\epsilon_d=
-0.025$, and $\Gamma=0.01$. When $\Delta=0$, the system goes from the 
free-orbital regime through the mixed valence regime directly to the frozen 
moment regime, without going through the local moment regime. The results
for the susceptibility, $<n_d>$, and $T<<S_z; S_z>>$ are shown in 
Fig.~\ref{mv}(a)-(c). Again, depending on the value of the band gap,
$T\chi$ can go to zero or 1/4 (there is a sharp transition).
For the cases that $T\chi$ goes to zero, $T<<S_z; S_z>>$ 
also goes to zero, and all correlation functions approach constants, 
which are independent of the band gap. But for the 
cases that $T\chi$ goes to 1/4, both $T<<S_z; S_z>>$ and 
the correlation functions approach the values which are 
band-gap dependent. 

\section{Effective Hamiltonian Description}

In this section we provide a simple interpretation of the low-temperature
behavior of the models in the various regimes on the basis of a simple
effective Hamiltonian. Let us consider first the Kondo Model with the gap in
the conduction electron density of states between $-\Delta$ to $\Delta$. 
The initial couplings are $J_0$
and $K_0$ in units of the bandwidth $D_0$, which is taken to be unity. In our
RG calculation, the band gap is taken to be $\Delta =\Lambda^{-M_0}$, where
$M_0$ is an integer; this corresponds to the maximum $N$ being 
$N_0 = 2M_0-1$ --- there are even number of conduction electron levels
in the discretized system. 

Imagine that we have successively integrated out the high energy degrees
of freedom and arrived at the effective Hamiltonian at the energy scale
$\Delta\equiv \Lambda^{-(N_0+1)/2}$; as we pointed out earlier the iterative 
RG procedure cannot be carried 
beyond this energy scale corresponding to the maximum 
iteration number $N_0$ since there are no conduction electron states left. 
The low temperature properties 
(i.e, for $T<<\Delta$) can be calculated with this effective Hamiltonian. 

Let us consider the case when $\Delta >> T_K$. The effective Hamiltonian
is close to that of the $J=0$ fixed point and can be written, keeping the
leading order terms, as
$$ H_{eff}=-J\vec{S}\cdot\vec{\sigma}(0)+Kf^{+}_{\mu}f_{\mu}
	+\Delta(a^{+}_{\mu}a_{\mu}-b^{+}_{\mu}b_{\mu})~~.$$
Here $\vec{\sigma}(0)=\frac{1}{2}f^+_{\mu}\vec{\sigma}_{\mu\nu}f_{\nu}$
and $f_{\mu} = \frac{1}{\sqrt{2}}(a_{\mu}+b_{\mu})$. In the above
effective Hamiltonian we have only kept the lowest single electron/hole
levels of the conduction electron Hamiltonian; these are
 represented by the creation operators $a^{+}_{\mu}$ and
$b^{+}_{\mu}$ and we have neglected the irrelevant operators. 
Since $f_0$ is proportional to $\Lambda^{-N_0/4}$, we have \cite{kgw} $$f_{0\mu} = \alpha_0 \Lambda^{-N_0/4}(a_{\mu}+b_{\mu})+....$$
Thus the first two terms of $H_{eff}$ are marginal, and $J$ and $K$ 
must scale as $J= J_0\Delta$ and $K=K_0\Delta$. When $|J_0|<<1$ and 
$|K_0|<<1$, the last term dominates, and the ground state is a doublet.

Next we consider the case $\Delta << T_K$. 
As we lower the energy scale to $\Delta$, the operator $f_0$ or $f$ is 
frozen, but $f_1$ is proportional to $\Lambda^{-N_0/4}$:
$$ f_{1\mu} = \hat{\alpha}_0\Lambda^{-N_0/4}g_{\mu}+....$$
The operator $g$ represents the single electron level at 
zero energy (the number of electron levels is odd, since $f_0$ is frozen). 

Now the effective Hamiltonian (at the energy scale $\Delta$) can be 
written as
\begin{equation}
\label{heff}
H_{eff}=-J\vec{S}\cdot\vec{\sigma}(0)+Kf^{+}_{\mu}f_{\mu}
	+w(f^{+}g+h.c.)~~.
\end{equation}
The operators $f$ and $g$ 
arise when we express $f_0$ and $f_1$ in terms of the lowest single
electron/hole levels of the conduction electron Hamiltonian. 
In the effective Hamiltonian given above 
$J$ and $K$ are renormalized coupling constants; they increase in 
magnitude as the high-energy degrees of freedom are integrated out but 
they saturate at the value 
attained at an energy scale of $T_K$ 
and are not further 
altered (since $f$ is frozen); however, the coupling constant
 $w$ will continue to scale as
$w \propto \Lambda^{-N_0/4}$ when $N_0$ increases (or as $\Delta$ 
decreases), we expect $w \propto \sqrt{\Delta}$. 
Since $w$ should be of the order of $T_K$ when $\Delta = T_K$, we can
re-write $w = \alpha T_K\sqrt{\Delta/T_K)}$. Note that in writing down the 
above Hamiltonian we have neglected all irrelevant terms, the inclusion of 
which will not change the results qualitatively.

We want to investigate the nature of the ground states of the above 
Hamiltonian for the cases $K_0=0$ and $K_0\neq 0$, by diagonalizing
the Hamiltonian.  This is mildly tedious but can be carried out in a
straightforward fashion. The main results are as follows:
The ground state is always a singlet (when $K>0$,
the ground state is in the subspace $(Q=-1,S=0)$ (for $K<0$, it 
is in the subspace $(1,0)$). The first excited
state is in the subspace (0,1/2) and has a gap relative to the ground state 
proportional to $\Delta$. 
For $K=0$, the ground state is in the subspace $(0,1/2)$, which is
a doublet with the energy gap
to the first excited state proportional to $\Delta^2$. These results are 
in agreement with our numerical RG computations. For the benefit of the 
reader a derivation of these results is presented below.

\subsection{Diagonalizing the effective Hamiltonian}

We diagonalize $H_{eff}$ in Equation (\ref{heff}) in two steps. 
Diagonalizing the first two terms of the Hamiltonian $H_{eff}$ in the 
subspace of the $f$ states 
gives rise to four eigenstates given below:
\begin{itemize}
\item State A $(-1,1/2)$: $E=0$
\item State B $(0,0)$: ~~$E=\frac{3}{4}J+K$
\item State C $(0,1)$: ~~$E=-\frac{1}{4}J+K$
\item State D $(1,1/2)$: ~$E=2K$
\end{itemize}
Here the numbers in the parentheses denote the charge and spin of the energy
states.

Now we add the $g$ states. The Hamiltonian can be written in the
basis consisting of A, B, C, D and $g$ states using a procedure similar
to what was employed 
in the iteration scheme of Wilson's RG iteration (see for example,
Eqn. (B2) in Appendix B of the paper by Krishna-murthy et al \cite{kww}). Let A1, A2, A3, A4 denote 
the basis states obtained by combining A with zero, one, and two $g$ states, 
etc. 
The Hamiltonian matrix in each charge-spin subspace can be written 
down as
\begin{itemize}
\item State A1 $(2, 1/2)$: $H_{2,1/2}=0$
\item State A3+B1 $(-1,0)$:
	$$H_{-1,0} = \left(\begin{array}{cc}
			0 & w \\
			w & \frac{3}{4}J+K \end{array}\right) $$
\item State A2+C1 $(-1,1)$:
	$$H_{-1,1} = \left(\begin{array}{cc}
			0 & w \\
			w & -\frac{1}{4}J+K \end{array}\right) $$
\item State A4+B2+C3+D1 $(0,1/2)$:
	$$H_{0,1/2}=\left(\begin{array}{cccc}
	0 & \frac{w}{\sqrt{2}} & -\sqrt{\frac{3}{2}}w & 0\\
	\frac{w}{\sqrt{2}} & \frac{3}{4}J+K & 0 &\frac{w}{\sqrt{2}}\\
	-\sqrt{\frac{3}{2}}w & 0 & -\frac{1}{4}J+K & 
	-\sqrt{\frac{3}{2}}w\\
	0 & \frac{w}{\sqrt{2}} & -\sqrt{\frac{3}{2}}w & 2K
	\end{array}\right)$$
\item State C2 $(0,3/2)$
	$$H_{0,3/2}=-\frac{1}{4}J+K$$
\item State B4+D3 $(1,0)$:
	$$H_{1,0} = \left(\begin{array}{cc}
			\frac{3}{4}J+K & -w \\
			-w & 2K \end{array}\right) $$
\item State C4+D2 $(1,1)$:
	$$H_{1,0} = \left(\begin{array}{cc}
			-\frac{1}{4}J+K & -w \\
			-w & 2K \end{array}\right) $$
\item State D4 $(2,1/2)$:
	$$H_{2,1/2} = 2K$$
\end{itemize}

Whether the ground state is a singlet or doublet 
depends on the relative energies of the lowest energy levels in subspaces 
$(-1,0),(1,0)$, and 
$(0,1/2)$. If the lowest energy level in the subspaces 
$(-1,0)$ and $(1,0)$ is lower than the lowest
energy level in the subspace $(0,1/2)$, then we have a singlet 
($T\chi$ will approach zero); otherwise, we have a doublet and 
$T\chi$ approaches $1/4$.

Let us first consider the case $K_0 \neq 0$. We perform
a second-order perturbation calculation of the energy of the eigenstate 
with the eigenvalue near $\frac{3}{4}J+K$:
\begin{itemize}
\item For the subspace $(-1,0)$, we have 
	$$E_0\approx \frac{3}{4}J+K+\frac{w^2}{\frac{3}{4}J+K}$$
\item For the subspace $(1,0)$, we have
	$$E_0 \approx \frac{3}{4}J+K+\frac{w^2}{\frac{3}{4}J-K}$$
\item For the subspace $(0,1/2)$, we have
	$$E_0 \approx \frac{3}{4}J+K+\frac{1}{2}\frac{w^2}{\frac{3}{4}J+K}
		+\frac{1}{2}\frac{w^2}{\frac{3}{4}J-K}$$
\end{itemize}

It is clear that the ground state is always a singlet: when $K>0$,
the ground state is in the subspace $(-1,0)$, whereas for $K<0$, 
the ground state
is in the subspace $(1,0)$). The energy level of the first excited
state (in subspace $(0,1/2)$) relative to the ground state is 
(assuming $K_0>0$)
$$E_1 \approx \frac{1}{2}\frac{w^2}{\frac{3}{4}J-K}
		-\frac{1}{2}\frac{w^2}{\frac{3}{4}J+K},$$
which is proportional to $\Delta$. The energy level of the second
excited state (in subspace $(Q=1,S=0)$) is $E_2 \approx 2E_1$ 
(this result was also found in our numerical results for the energy levels).

How about $K_0 = 0$? The issue cannot be resolved at the level of 
second-order perturbation theory. A fourth-order 
perturbation calculation for the lowest energy in the subspace
$(0,1/2)$ yields 
$$E_0 \approx
	\frac{3}{4}J+\frac{4}{3}\frac{w^2}{J}+\frac{80}{27}J
	(\frac{w}{J})^4.$$
For the subspaces $(-1,0)$ and $(1,0)$, the lowest energy is given by
$$E_0 = \frac{3}{4}J-\sqrt{(\frac{3}{4}J)^2+w^2}\approx
	\frac{3}{4}J+\frac{4}{3}\frac{w^2}{J}-\frac{1}{2}J
	(\frac{w}{J})^4.$$

It is clear that the ground state is in the subspace $(0,1/2)$, which is
a doublet. This agrees with our numerical results. The energy gap
of the first excited state is proportional to $w^4$ or
$\Delta^2$.

\subsection{Response Function}

Let us consider the calculation of $T<<S_z; S_z>>$ in the ground state 
when $\Delta  << T_K$. 
By definition
	$$<<S_z; S_z>>=\int^{\beta}_0 <S_z(\tau)S_z>d\tau.$$
Close to zero temperature, we can write 
$$<<S_z; S_z>>=\sum_{|I>}\frac{|<G|S_z|I>|^2(1-\exp(-\beta(E_I-E_G)))}
	{E_I-E_G},$$
where $|I>$ represents many-body states of the system and $|G>$
denotes the ground state. For temperatures much smaller 
than the energy gap between the first excited state and the ground state, we
have (separating out the contribution of the ground state from the
summation)
$$<<S_z; S_z>>=\beta |<G|S_z|G>|^2+
	\sum_{|I>\neq |G>}\frac{|<G|S_z|I>|^2}
	{E_I-E_G}.$$
Since the second term in the above expression is finite, we obtain in 
the limit 
as $T\rightarrow 0$, 
$$T<<S_z; S_z>>=|<G|S_z|G>|^2 ~.$$
For the case $K_0 \neq 0$, it is easy to verify that 
$T<<S_z; S_z>>$ is zero.

For the case that $K_0=0$, 
we find that 
$$<G|S_z|G>=\pm \frac{4w^2}{3J^2}.$$
Thus $T<<S_z; S_z>>$ in this case is proportional to $w^4$ or
$\Delta^2$ in agreement with the numerical results. 

\subsection{Anderson Model}

We now discuss the Anderson Model briefly since the results are similar 
to those of the Kondo problem discussed above. We consider the limit that 
U is very large and $\Gamma$ is very small. The effective Hamiltonian 
is of the form
$$H_{eff} = \epsilon_d d^{+}_{\mu}d_{\mu}
	+V(d^{+}_{\mu}f_{0\mu}+h.c.)+\Delta(a^{+}_{\mu}a_{\mu}
		-b^{+}_{\mu}b_{\mu}). $$
Here $\epsilon_d$ is the effective impurity level at the energy scale 
$\Delta$ and $V$ is the effective coupling to the conduction electron
states. The impurity level cannot be doubly occupied: $n_d\leq 1$.
For the case that $-\epsilon_d>\Delta$, then the local moment
regime will be reached, and the effective Hamiltonian can be converted 
to the Kondo Hamiltonian; this has been discussed above. Here we focus
on the case that $-\epsilon_d << \Delta$.

Consider the case when $V$ is very small; to leading
order, the ground state depends on the sign of $\epsilon_d$.
If $\epsilon_d>0$, then the ground state corresponds to two electrons
occupying the conduction electron level at $-\Delta$, and it is a singlet 
(this 
situation arises, for example, when the initial $\epsilon_d$ is greater 
than zero; this has been checked by the nonperturbative RG calculation).
If $\epsilon_d<0$, then the ground state corresponds to
two electrons occupying the conduction electron level at $-\Delta$ and
one electron occupying the impurity level. So the ground state is a
doublet. 

From the above analysis, it is clear that the reason for the ground
state being a doublet for the case $-\epsilon_d> \Delta$ (when the local
moment regime is reached) and $-\epsilon_d <<\Delta$ (when only the 
mixed-valent
regime is reached) are different. In the first case, the local
moment regime is reached, and as the temperature is lowered, the moment 
begins to be 
quenched due to large effective $|J|$, but as the temperature is further
lowered, one can see the small splitting of the singlet state to the doublet
state due to the finite gap energy. In the second case, the local moment 
is not formed at the energy scale $\Delta$. But as the temperature is 
lowered to the energy scale of $-\epsilon_d$, charge fluctuations are 
suppressed and they eventually 
cease to exist, and the system becomes a doublet.

\section{Conclusions}

We have performed a Wilson Renormalization group calculation of the 
Kondo and Anderson models with a gap in the conduction electron 
density of states. The impurity
susceptibility, correlation functions, and a zero-frequency response
function have been calculated as functions of temperatures in various 
regimes. Our calculations confirm earlier results on the
 qualitative differences in the low-temperature behaviors
 between the cases with and without particle-hole symmetry
when the gap is much smaller than the Kondo temperature. We have shown
that the numerical results at low temperatures can be understood 
in terms of simple low temperature effective Hamiltonians. \\

{\em Acknowledgments} C.J would like to thank the Ohio Supercomputer Center
for granting computer time on the Cray.

\begin{table}
\caption{The values of $T<<S_z; S_z>>$ at zero temperature for
a range of values of the band gap. The numbers enclosed in the parentheses
are the total charge $Q$ and spin $S$ of the ground state.
\label{c0}}
\begin{tabular}{|c|c|c|}
$\Delta$ & $J_0=-0.2,K=0.0$ & $J_0=-0.2,K=0.0$\\ \tableline
$1.88\times 10^{-6}$ & 0.0398 (0, 1/2) & 0.0490 (0, 1/2) \\ \tableline
$6.27\times 10^{-7}$ & 0.00797 (0, 1/2) & 0.0110 (0, 1/2) \\ \tableline
$2.09\times 10^{-7}$ & 0.00101 (0, 1/2) & $<1.2\times 10^{-5}$ ($-1$, 0)
\\ \tableline
$6.96\times 10^{-8}$ & 0.000115 (0, 1/2) & $<4.0\times 10^{-6}$ ($-1$, 0)
\\ 
\end{tabular}
\end{table}
 

\begin{references}


\bibitem{kgw}  K. G. Wilson,  {\em  Rev. Mod. Phys.} {\bf 47}, 773 (1975).

\bibitem{nozieres} P. Nozi\`{e}res, {\em J. Low Temp. Phys.} {\bf 17}, 31 (1974).

\bibitem{kww} H. R. Krishna-murthy, J. W. Wilkins, and K. G. Wilson,
{\em Phys. Rev. B} {\bf 21}, 1003 and 1044 (1980).

\bibitem{cruz} L. Cruz, P. Phillips, and A. H. Castro Neto, Europhys. Lett.,
{\bf 29}, 389 (1995).

\bibitem{taka} H. Takayama, Y. R. Lin-Liu, and K. Maki, Phys. Rev. B {\bf 21},
2388 (1980).

\bibitem{cj1} K. Chen and C. Jayaprakash,  Phys. Rev. B {\bf 52}, 14436.
(1995). 

\bibitem{cjk} Kan Chen, C. Jayaprakash, and H.R. Krishnamurthy, 
	{\em Phys. Rev. B} {\bf 45}, 5368-5386 (1992).

\bibitem{cj2} Kan Chen and C. Jayaprakash, J. Phys: Condens. Matter {\bf 7}, 
L491 (1995). 

\bibitem{saso0} T. Saso, {\em J. Phys. Soc.
Jpn.} {\bf 61}, 3439 (1992)

\bibitem{saso1} T. Saso  and J. Ogura, Physica {\bf B~186-188}, 372 (1993)

\bibitem{ogura1} J. Ogura and T. Saso, {\em J. Phys. Soc.
Jpn.} {\bf 62}, 4364 (1993)

\bibitem{take1} K. Takegahara, Y. Shimizu, and O. Sakai, {\em J. Phys. Soc.
Jpn.} {\bf 61}, 3443 (1992)

\bibitem{take2} K. Takegahara, Y. Shimizu, N. Goto, and O. Sakai,
Physica {\bf B~186-188}, 381 (1993)

%\bibitem{satori} K. Satori, H. Shiba, O. Sakai, and Y. Shimizu, {\em J. Phys.
%Soc. Jpn.} {\bf 61}, 3239 (1992)

\bibitem{yu} C. C. Yu and M. Guerrero, {\em Phys. Rev.} {\bf B 54}, 8556
(1996)

\bibitem{inger} K. Ingersent, {\em Phys. Rev.} {\bf B 54}, 11936 (1996)

%\bibitem{inger2} C. Gonzalez-Buxton and K. Ingersent, {\em Phys. Rev.} 
%{\bf B 54}, R15614  (1997)

\end{references}
\end{document}